\documentclass[11pt]{article}
\usepackage{epsfig}
\def\Journal#1#2#3#4{{#1} {\bf #2}, #3 (#4)}

\def\NPB{{\em Nucl. Phys.} B}
\def\PLB{{\em Phys. Lett.}  B}
\def\PRL{\em Phys. Rev. Lett.}
\def\PRD{{\em Phys. Rev.} D}
\def\ZPC{{\em Z. Phys.} C}
\def\EJC{{\em Eur. Phys. Jour.} C}
\def\SJNP{\em Sov. J. Nuc. Phys.}
\def\rs{\sqrt s}
\def\ms{\mbox{
   $\Lambda^{{\scriptscriptstyle (4)}}_
   {\overline{{\scriptscriptstyle MS}}}$}}
\begin{document}
\rightline{LAPTH-Conf-734/99}
\vspace{4 cm}
\begin{center}
{\Large {\bf{INCLUSIVE PHOTON PRODUCTION IN HADRONIC COLLISIONS\\}}}
\vspace{0.7 cm}
{\large M.~WERLEN }

Laboratoire d'Annecy-le-Vieux de Physique Th\'eorique LAPTH
\footnote{URA 1436 du
CNRS, associ\'ee \`a l'Universit\'e de Savoie.},\\
LAPP, B.P. 110, F-74941 Annecy-le-Vieux Cedex, France\\
and EP division, CERN, CH-1211 Geneva 23
\end{center}

\centerline{\bf{Abstract}}
   High statistics fixed target and ISR inclusive photon production 
   data are compared to next-to-leading order (NLO) QCD calculations.
   The dependence of the theoretical predictions on the structure functions, 
   and on the theoretical scales is investigated.
   It is shown that the data cannot be simultaneously fitted
   with a single set of structure functions and scales. However,
   it is argued that as long as one restricts 
   the data/theory comparison to the $x_T$ range where the theory is reliable,
   i.e. stable with respect to the scale variation,
   there is no need to introduce  an additional primordial $k_{_T}$
   dependence except for Be data. Finally a precise determination 
   of the strong coupling constant, $\alpha_s$, is performed from the
   direct photon production cross sections obtained in high statistics
   $\bar{p}p$ and $pp$ collisions  at the CERN SPS
   (UA6) by a NLO QCD analysis.
\vskip 4cm
\begin{center}   
Talk at the 34th Rencontre de Moriond, \\
QCD and High Energy Hadronic Interactions\\
 Les Arcs, France, 20 - 27 Mar 1999.
\end{center}
\newpage
\section{Critical study of photon production in pp and pBe interactions}
Inclusive photon production 
including the direct and the bremsstrahlung contributions, 
calculated at NLO in QCD \cite{critic}, 
is compared to experimental data in the range $\rs$=23 GeV to 63 GeV
as a function of $x_T=2 p_T/\sqrt{s}$. The following experimental data
are used: the $pp$ data from WA70~\cite{WA70} and from ISR
R806~\cite{R806}, R110~\cite{R110}, AFS~\cite{AFS}
and R108 \cite{R108} as well as the new fixed target data from E706~\cite{E706}
($pBe$) and UA6~\cite{UA6} ($pp$ and $p\bar{p}$).
The main results of this study are summarized below. 
They concern the theoretical prediction uncertainty and possible incompatibilities
between data sets.

The perturbative NLO calculations involve three arbitrary 
scales: the factorization scale $M$, the renormalization scale $\mu$
and the fragmentation scale $M_F$. As those scales are unphysical,
the theory can be considered reliable only in the region of the phase space
where the predictions are stable with respect to the scale variations.
For a large range of $M_F$ values, the theory stability versus 
$M$ and $\mu$ is reasonable for  $x_{_T}> .35$
at $\rs$=23 GeV, $x_{_T}> .32$ at $\rs$=31.6 GeV,  $x_{_T}> .26$
at $\rs$=38.8 GeV and $x_{_T}> .16$ at $\rs$=63 GeV.
The theoretical uncertainty due to the scales is approximated by
the difference in the predictions between $M=\mu=M_F=p_T/3$ and $M=\mu=M_F=p_T/2$.

\begin{figure}[htb]
\vspace{-1.0cm}
\begin{center}
\epsfig{file=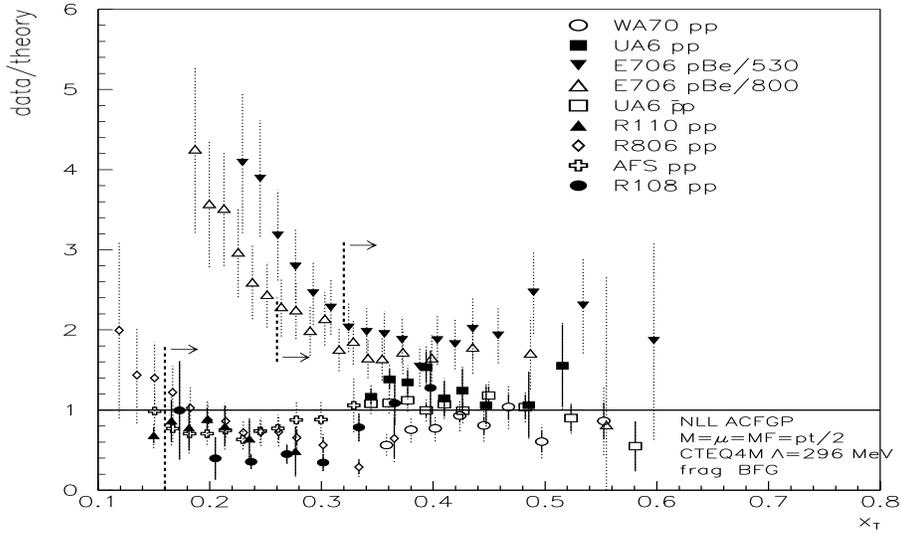,width=13cm,height=8cm}
\end{center}
\vspace{-0.5cm}
\caption{Ratio data/theory for direct photon production using CTEQ4M
distribution functions. All scales are set to $p_T/2$.
The arrows note the $x_T$ ranges where the perturbative predictions
are reasonably stable vs variation of the scales $\mu$ and $M$. Statistical errors are shown as
full lines, statistical and systematic errors added in quadrature
are shown as dashed lines. }
\label{cteq4m}
\end{figure}

\begin{figure}[htb]
\begin{center}
\epsfig{file=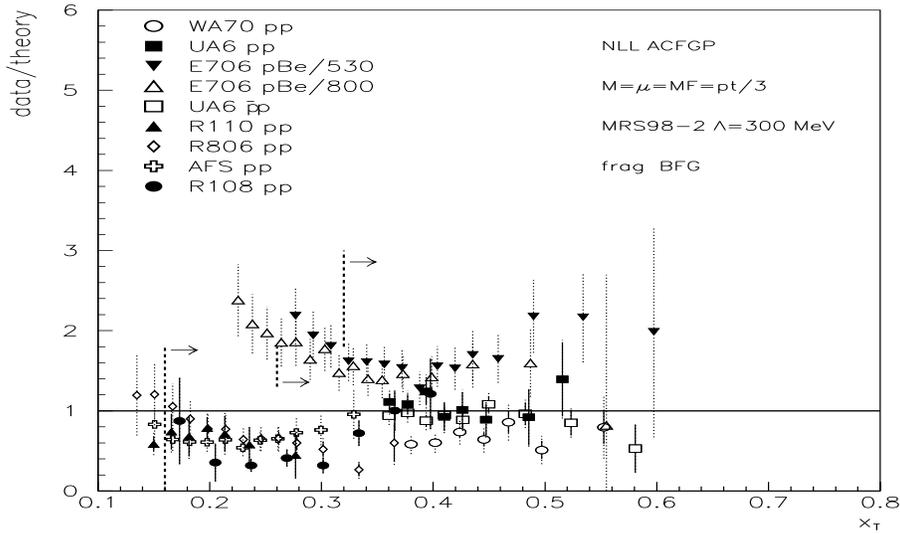,width=13cm,height=8cm}
\end{center}
\caption{Ratio data/theory for direct photon production using MRS98-2
distribution functions. All scales are $p_T/3$.
}
\label{mrs98}
\end{figure}

The ratio data/theory is shown in Fig.~\ref{cteq4m}
with all scales set to $p_T/2$ and using the parton distribution 
functions CTEQ4M \cite{cteq4}.
A large discrepancy can be seen 
between the E706 and the other data sets. Although there is an approximate
agreement
with theory for the WA70, UA6 and ISR data sets,  the E706 data
are underestimated by the theory by a factor 2 in the region 
of theory stability and up to 4 at low $x_T$. 
In Fig.~\ref{mrs98} the same experimental data are compared with the theory
with all scales set to $p_T/3$ and using  the parton distribution functions
MRS98-2~\cite{mrs98}. Here too the theory underestimates
the E706 data by 50\% in the region of theory stability
and by up to 100\% at low $x_T$.

Clearly, the full set of available data cannot be fitted by
the theory whatever the scales or the structure functions.
Restricting the comparison to the $x_T$ range of theory stability
reduces the discrepancy to a mere global scaling of the E706 data.

It has been proposed to introduce a  parton-$k_T$ broadening effect
\cite{E706,hus} to account for the low $x_T$ dependence and for the 
scaling effect seen by E706. This effect would destroy the theoretical agreement
of all other experiments particularly in $x_T$ region of theory stability.

\begin{figure}[htb]
\begin{center}
\epsfig{figure=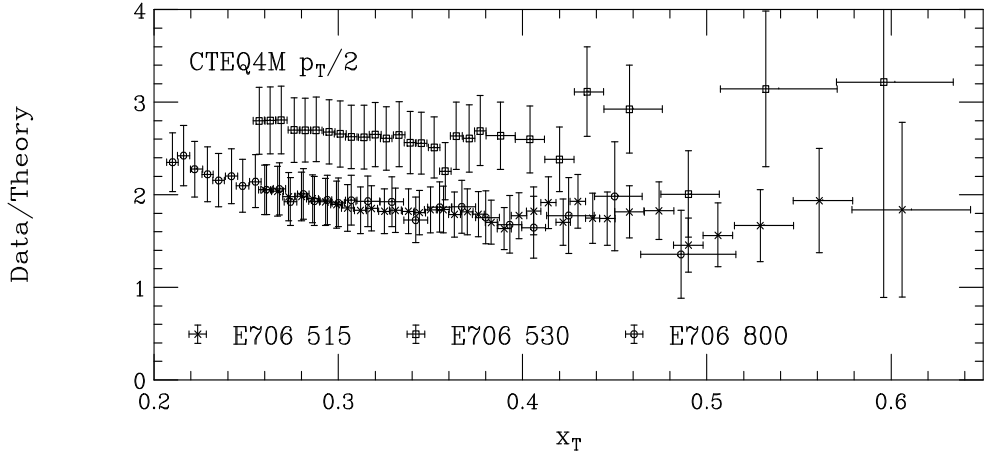,width=6cm,height=13cm,angle=90}
\end{center}
\caption{Ratio data/theory for $\pi^0$  production using CTEQ4M
parton distribution functions. All scales are set to $p_T/2$.
The E706 pBe data with beams of 530 GeV/c and 800 GeV/c
as well as the E706 $\pi^-$Be data with a beam of
515 GeV/c (not discussed here) are shown.}
\label{piz}
\end{figure}

A NLO QCD calculation has been performed 
for $\pi^0$ production \cite{pizero}. Setting all scales to $p_T/2$, 
the low $x_T$ rise observed in the direct photon case (Fig.~\ref{cteq4m}) 
has almost disappeared (Fig.\ref{piz}). Moreover, when all scales are set to 
$p_T/3$, a common rescaling of the theory by a factor $1.5\pm0.2$
is enough to accommodate all fixed target data. This normalization
may not be unrealistic as
the absolute normalization of the theory is not precise,
due for example to present uncertainties in the $\pi$ fragmentation.
The fixed target $\pi^0$ data appear in better agreement
then the photon data.

\section{Determination of $\alpha_s$ from UA6 data ~\protect\cite{alphas}}
\begin{figure}[htb]
\begin{center}
\begin{tabular}{lr}
\epsfig{file=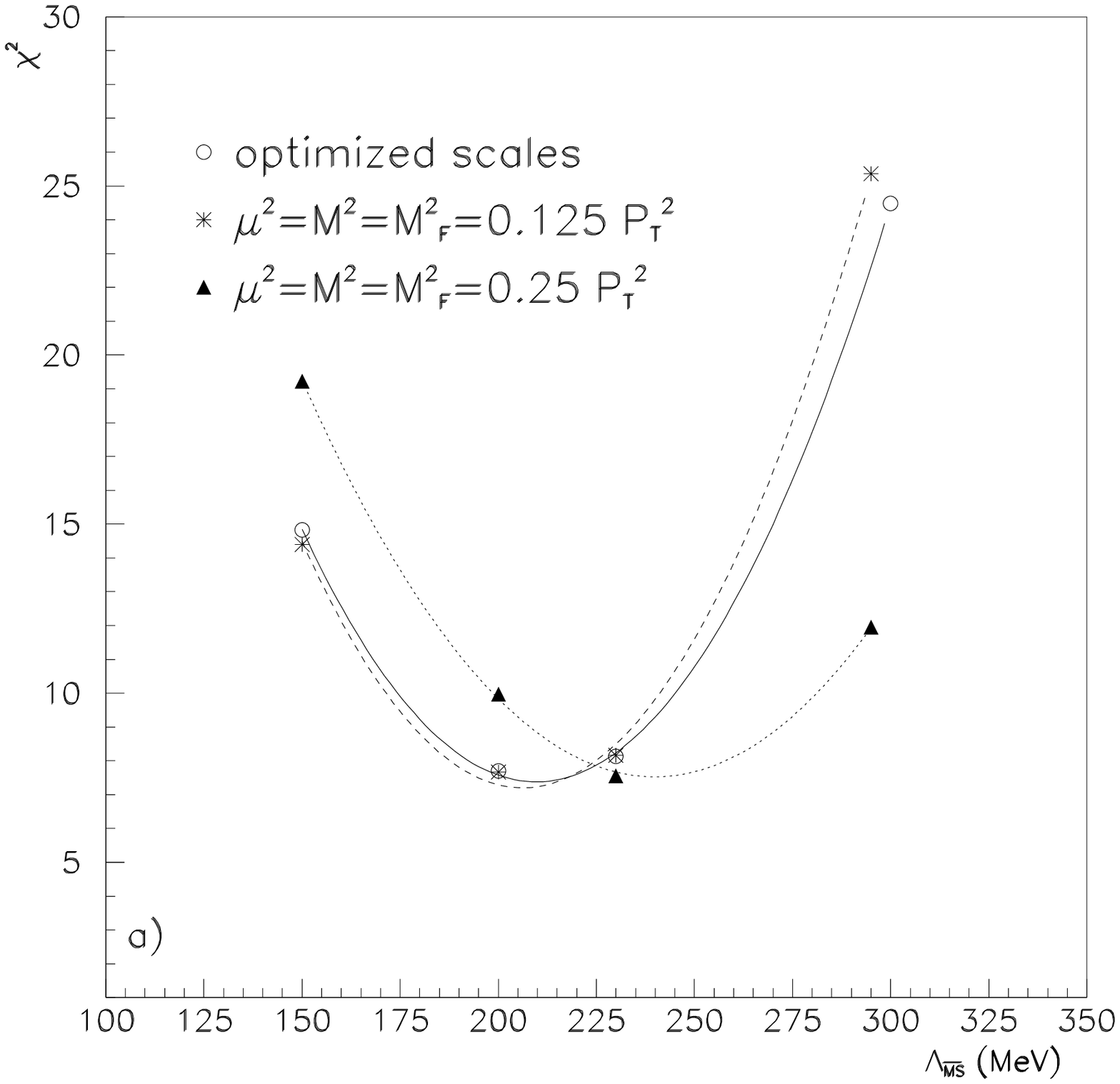,width=7.cm,height=7.cm} &  
\epsfig{file=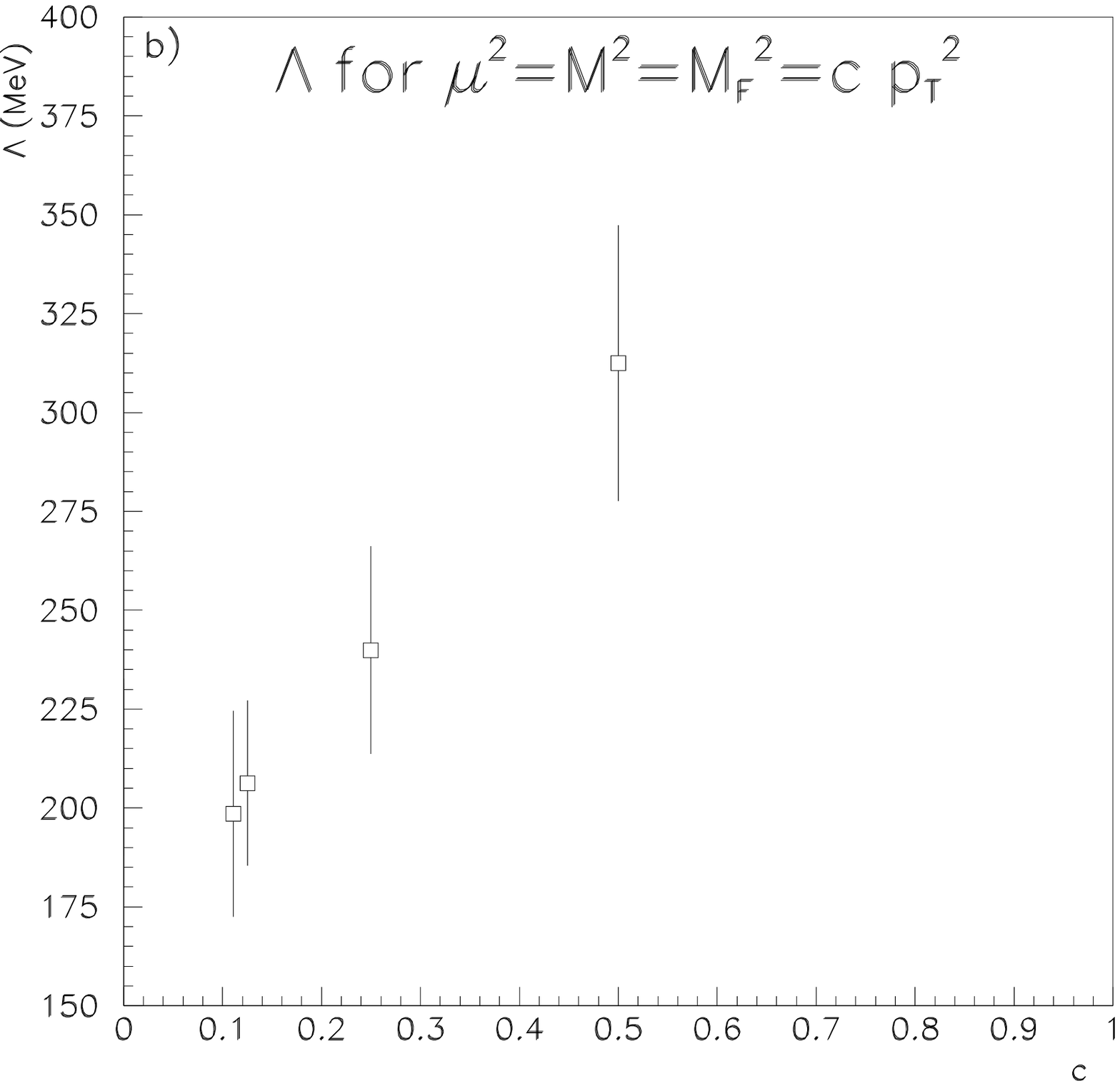,width=7.cm,height=7.cm} \\
\end{tabular}
\end{center}
\caption{a) The $\chi^2$ between the theoretical predictions and the
measured cross section difference 
$\frac{d\sigma(\bar{p}p \to \gamma X)}{dp_{Ti}} - \frac{d\sigma(pp \to \gamma X)}
{dp_{Ti}} $ summed
over ten $p_{Ti}$ bins
as a function of \ms\ for various choices of scales.
b) Best value of \ms\ as a function of the parameter c defining the
 scales $\mu^2 = M^2 = M_F^2=c p_T^2$. The error bars are statistical only.}
\label{lambda}
\end{figure}

The measurement in the same experiment of the production of direct
photon in $pp$ and $\bar{p}p$ collisions \cite{UA6} allows a clean isolation
of the annihilation process ($q \bar{q} \to \gamma g$) from the difference
$\sigma(\bar{p}p \to \gamma X)$ $ - \sigma(pp \to \gamma X)$.
Knowing the quark distributions, this process provides a direct handle
on \ms\  and therefore $\alpha_s$.
Quark distributions vs \ms\  have been  obtained 
from fits to BCDMS \cite{bcdms} deep
inelastic scattering data. Therefore \ms\  remains the only 
free  parameter in the fit of the  UA6 data
to the  NLO theoretical calculations. The results are shown
in Fig. \ref{lambda} for various choices of scales. 

For optimized scales \cite{opt}, the best $\chi^2$  gives
\ms\ of
$210 \pm 22 \,(stat.) \pm 44 \,(syst.) $ MeV.
We conservatively assign an overall theoretical uncertainty of $^{+105}_{-36}$ MeV,
dominated by uncertainty in the $\mu$ and M scale selection.

This leads to 
$\alpha_s(M^2_Z)=0.1112~\pm 0.0016 \,(stat.)
~\pm 0.0033 \,(syst.)~^{+0.0077}_{-0.0034} \,(theo.)$
in good agreement with the world average value \cite{summer98}
and with  error bars comparable to those achieved in deep
inelastic scattering \cite{vm,ccfr}.
Recent improved NLO computation for $\bar{p}p$
including resummation shows small corrections for small scales
as used here and furthermore reduces the scale dependence 
as expected \cite{resum}.
Hence these new calculations should not change significantly
the $\alpha_s$ value.

\section*{References}
\begin{thebibliography}{99}

\bibitem{critic} P.~Aurenche, M.~Fontannaz, J.Ph.~Guillet,  B.~Kniehl, E.~Pilon,
M.~Werlen, hep-ph/9811382.  


\bibitem{WA70} WA70 Collaboration, M.~Bonesini {\it et al}, \Journal{\ZPC}{38}{535}{1988}.
     
\bibitem{R806} R806 Collaboration, E.~Annassontzis {\it et al}, \Journal{\ZPC}{13}{277}{1982}. 
     
\bibitem{R110}  R110 Collaboration, A.L.S.~Angelis {\it et al}, \Journal{\NPB}{327}{541}{1989}.
     
\bibitem{AFS}  AFS/R807 Collaboration, T.~\AA kesson {\it et al}, \Journal{\SJNP}{51}{836}{1990}.
      
\bibitem{R108} R108 collaboration, A.L.S.~Angelis {\it et al}, \Journal{\PLB}{94}{106}{1980}. 
	
\bibitem{E706} E706 Collaboration, L.~Apanasevich {\it et al}, \Journal{\PRL}{81}{2642}{1999}. 
 	        
\bibitem{UA6}  UA6 Collaboration, G.~Ballocchi {\it et al}, \Journal{\PLB}{436}{222}{1998}.
    
\bibitem{cteq4} H.L.~Lai {\it et al}, \Journal{\PRD}{55}{1280}{1997}. 
    
\bibitem{mrs98} A.D.~Martin, R.G.~Roberts, W.J.~Stirling, R.S~Thorne, \Journal{\EJC}{2}{529}{1998}.
    
\bibitem{hus} L.~Apanasevich {\it et al}, \Journal{\PRD}{59}{074007}{1999}. 
   
\bibitem{pizero} P.~Aurenche, M.~Fontannaz, J.Ph.~Guillet, B.~Kniehl, E.~Pilon,
M.~Werlen, in preparation.  

\bibitem{alphas} UA6 collaboration, M. Werlen {\it et al}, \Journal{\PLB}{452}{201}{1999}.
 
\bibitem{bcdms} BCDMS Collaboration, A.C.~Benvenuti {\it et al}, \Journal{\PLB}{223}{485 and 490}{1989};
\Journal{\PLB}{237}{592 and 599}{1990}. 

\bibitem{opt} P.~Aurenche, R.~Baier, M.~Fontannaz and D.~Schiff,
\Journal{\NPB}{286}{509}{1987}.

\bibitem{summer98} Y.L.~Dokshitzer, ICHEP-98 Vancouver, hep-ph/9812252.

\bibitem{vm} M.~Virchaux and A.~Milsztajn, \Journal{\PLB}{274}{221}{1992}. 

\bibitem{ccfr} CCFR Collaboration, W.G.~Seligman {\it et al}, \Journal{\PRL}{79}{1213}{1997}.

\bibitem{resum} S.~Catani, M.~Mangano, P.~Nason, C.~Oleari, W.~Vogelsang, hep-ph/990346. 


\end{thebibliography}
\end{document}